 \documentclass[final,5p,times,twocolumn,authoryear]{elsarticle}

\usepackage{amssymb}
\usepackage{lipsum}
\usepackage{multirow}
\usepackage{graphicx}
\usepackage{siunitx}
\PassOptionsToPackage{hyphens}{url}\usepackage{hyperref}
\usepackage{cleveref}
\usepackage[utf8]{inputenc}
\usepackage[right]{lineno}
\usepackage[table,xcdraw]{xcolor}
\usepackage{csquotes}
\usepackage{booktabs}
\usepackage{longtable}
\usepackage[dvipsnames]{xcolor}
\usepackage{adjustbox}
\usepackage{tikz}
\usetikzlibrary{arrows.meta,positioning,shadows.blur,fit}
\usepackage{array}
\usepackage{url}
\usepackage{titlesec}
\usepackage{graphicx}
\usepackage{subcaption}

\begin{document}

\begin{frontmatter}



\title{From Perception to Pixels: Human-Derived Visually Distinct Colors for Improved Dominant Color Extraction}


\author{Uldana Shyndali, Nurdaulet Toleugaliyev, Kuat Zharykbayev, Zarina Abildasheva, Pakizar Shamoi}

\affiliation{organization={Kazakh-British Technical University, School of Information Technology and Engineering},
            addressline={Tole bi 59}, 
            city={Almaty},
            postcode={050000}, 
            country={Kazakhstan}}

\begin{abstract}
Dominant color extraction is a fundamental task in computer vision, with applications in design, image retrieval, and content recommendation. However, conventional algorithms based on device-oriented color spaces (e.g., RGB) and fixed quantization often fail to capture how humans perceive color differences. In this study, we propose a human-perception-based framework for dominant color extraction, built on a novel Human Color Dictionary (HCD) derived from a perceptual experiment. Participants (n=110) were sequentially shown 120 colors and asked to identify those they perceived as unique. Based on their responses, we constructed a $120\times120$ co-occurrence matrix and applied a greedy selection algorithm to derive 64 perceptually distinct colors, forming the HCD. This dictionary was integrated into RGB color histogram and K-means color extraction algorithms by initializing centroids and mapping pixel values to their nearest perceptual color using the $\Delta E_{00}$ distance. Evaluation on a dataset of 4,562 images paired with expert-designed palettes demonstrated that our method improves perceptual alignment, as measured by Earth Mover’s Distance (EMD) and best-match distance. Our findings suggest no meaningful gender differences in the number of uniquely perceived colors, indicating that sex does not significantly influence color perception in this task. Another key finding is that highly saturated colors are perceived as the most distinguishable, whereas dark and deep colors tend to perceptually collapse and are less reliably differentiated. The results confirm that incorporating experimentally derived human color representations 
ensures the generated color clusters align more closely with human perception, thus improving the usability of color palette generation in practical applications. 
\end{abstract}





\begin{keyword}

Human perception \sep Color clustering \sep k-means clustering \sep  CIE Lab\sep  Histogram\sep  Dominant colors extraction



\end{keyword}

\end{frontmatter}




\section{Introduction}

Color plays a central role in how humans interpret visual information \cite{colibri}. From design and marketing to digital art and user interface aesthetics, the dominance and composition of colors influence perception, emotion, and decision-making \cite{11139293, 10597554}. When individuals assess something new, 62-90\% of their evaluation during the first 90 seconds is based on color, highlighting the significance of color in human perception \cite{Singh2006_ImpactColorMarketing_1}. 

Therefore, determining the dominant colors in an image is a key task in computer vision and visual computing applications, including image retrieval, automatic palette generation, visual data analysis in marketing and aesthetics, and content-based recommendation systems \cite{Kynash2025New, 9869653_4, Gunduz2021A, 10001872}. Despite the abundance of algorithmic approaches, achieving results that correspond to human perception remains a fundamental challenge \cite{Chang2022, Nadler2023Divergences}.

Most existing algorithms operate in device-oriented color spaces such as RGB and rely on purely statistical grouping of pixel values \cite{Burney2014K-Means, Jardim2022Graphical}. Dominant colors can also be extracted efficiently without full pixel-domain processing by working directly in the compressed (DCT) domain, which is useful for large-scale pipelines \cite{Jiang2006}. These approaches treat colors as numerical entities, neglecting how humans perceive color differences and boundaries. As a result, multiple visually similar shades are often identified as distinct clusters, while perceptually dominant hues may be underrepresented because they occupy smaller pixel areas. This discrepancy limits the interpretability and human relevance of color extraction outputs. Pixel-level descriptors may deviate from human visual judgments and can be disproportionately affected by small outlier regions, motivating perceptually grounded dominant-color representations \cite{Kiranyaz2010}.



One of the most commonly used methods for determining dominant colors is K-means clustering. This method groups image pixels based on their color characteristics, typically using RGB values \cite{Kanungo2002An}. Despite its widespread use, K-means has several challenges. It is sensitive to the initial choice of centroids, struggles to determine the optimal number of clusters, and relies on Euclidean distance in RGB space \cite{Basar2020Unsupervised, Burney2014K-Means}, which often does not reflect how people perceive color differences. In broader terms, dominant-color extraction is commonly framed as color quantization - a two-stage process of palette design and pixel mapping - which helps explain why these practical choices can strongly shape the final palette \cite{Celebi2011}. These limitations can lead to results that are computationally accurate but perceptionally contradictory.
\\


 To address this limitation, we introduce a human perception–based framework for dominant color extraction that integrates experimentally derived perceptual data into the computational process. The proposed approach is grounded in a color recognition experiment involving more than one hundred participants who evaluated 120 candidate colors. From these responses, a perceptually distinct set of 64 colors—termed the Human Color Dictionary (HCD) was constructed through matrix-based co-occurrence analysis and a greedy selection algorithm. 
 

Next, we integrate this dictionary into established algorithms, including K-means and RGB color histograms, by mapping pixel values to their nearest perceptually similar color using the $\Delta E_{00}$ distance in CIE~Lab space. The effectiveness of this integration is evaluated against a ground-truth dataset of 4,562 images with expert-designed palettes, using best-match distance and Earth Mover’s Distance (EMD) as perceptual similarity metrics.




The paper has the following contributions:
\begin{itemize}
    \item \textit{Human-Derived Color Dictionary.} We conducted a perceptual experiment with 110 participants, resulting in a 64-color Human Color Dictionary (HCD) that captures perceptually distinct hues recognizable by human observers
   
   \item \textit{Perceptual Integration} of this dictionary to traditional dominant color extraction algorithms, leading to improvement in alignment with human perception.
    \item \textit{Reproducible methodology} for constructing perceptual color dictionaries and a framework for embedding them into established extraction algorithms
   \item \textit{Ground-Truth Evaluation Framework.} A benchmark dataset of 4,562 images with expert-designed palettes was collected and used as perceptual ground truth for evaluation.
\end{itemize}



The paper is structured as follows. Section I is this Introduction, Section II reviews related work on clustering- and quantization-based approaches, perceptual color studies, and challenges in pixel-based methods. Section III details the collection of the dataset from expert-designed palettes, the baseline algorithms, and the proposed human-derived color dictionary from a perceptual experiment. Section IV explains the experimental setup using ground-truth palettes and perceptual similarity metrics. Section V presents results and application examples. Finally, Section VI concludes the paper and provides directions for future research.

\section{Related Work}

Color clustering and dominant color extraction have been extensively studied in image processing, computer vision, and design-related applications. Existing research focuses primarily on clustering-based techniques, perceptual color modeling, and application-driven adaptations. 


\subsection{Clustering-Based Approaches to Color Extraction}



Clustering algorithms, particularly K-means, are among the most widely used techniques for color segmentation and dominant color extraction because of their simplicity and computational efficiency. Early studies demonstrated the effectiveness of K-means for grouping color data, but also revealed its limitations in capturing perceptual similarity. Even when the algorithm is fixed, the resulting palette can vary substantially depending on how the procedure is configured and executed \cite{Celebi2011}. For example, \cite{GedeKarmaSudarma_ColorKMeans} clustered hundreds of color samples into predefined cluster sizes and observed that colors with small RGB differences were frequently grouped together, even when perceptually distinguishable. This highlights the inadequacy of RGB-based distance metrics for modeling human color perception.

Several studies explored enhancements to K-means through preprocessing or alternative feature representations. In satellite image segmentation, \cite{ChitadeKatiyar2010} applied decorrelation stretching prior to clustering to improve color separability and reduce computational cost. Similarly, \cite{Hussain2016ColorConstancy} employed K-means to segment uniform color regions for color constancy correction, demonstrating improved robustness under varying illumination. The influence of color space selection on clustering performance was systematically examined by \cite{Jurio}, who compared RGB, HSV, YUV, and CMY spaces, emphasizing that clustering accuracy is highly dependent on the underlying color representation.

Beyond segmentation, clustering has been applied to extract dominant colors for higher-level analysis. For example, \cite{10759889_11} used K-means to identify dominant logo colors and map them to emotional categories in branding studies. However, these approaches still rely on pixel-based statistics and predefined cluster numbers, often leading to over-segmentation or perceptually inconsistent results.

\subsection{Perceptual and Human-Centered Color Modeling}

To overcome the perceptual limitations of traditional clustering, several studies have incorporated perceptual color spaces and human-centered models. The use of CIE-based spaces, such as CIELAB and CIELUV, has been shown to better approximate human color discrimination. In practice, this is often done because converting RGB to CIE Lab gives perceptual significance to the color distances used in clustering decisions \cite{Pagnutti2018}. In this context, \cite{Douven2017ClusteringColors} investigated whether computational clustering can approximate human color lexicons, demonstrating that perceptually uniform spaces improve alignment with human-defined color categories.

Other works explicitly modeled perceptual attributes such as saturation, contrast, and saliency. \cite{Chang2022} proposed a dominant color extraction method that integrates these perceptual factors, showing that dominant color selection is influenced not only by pixel frequency but also by visual prominence. Similarly, \cite{Liu2021_BackgroundDominantColors_3} introduced a quick fuzzy C-means–based approach tailored for background color extraction, leveraging spatial and perceptual cues to improve clustering stability.

Foundational contributions in perceptual color representation include the MPEG-7 dominant color descriptor introduced by \cite{Manjunath2001}, which combines color histograms with perceptual relevance. Extensions toward perceptually meaningful palette analysis were further explored by \cite{Phan2018ColorOrchestra}, who applied density estimation to organize color palettes extracted from artworks, enabling applications such as color harmonization and style transfer. Additionally, \cite{6005985_14} proposed a color perception model for image retrieval that incorporates intensity and contrast, further demonstrating the importance of perceptual attributes in color-based analysis.

\subsection{Application-Driven, Interactive, and Fuzzy Extensions}

Recent research has increasingly focused on adapting color extraction techniques to specific application domains and interactive workflows. In document design analysis, \cite{Perera2023DominantPalette} introduced a color pixel quantifier algorithm to extract dominant palettes from resumes, prioritizing visual consistency over purely statistical dominance. In the context of creative tools, \cite{Shi2024} explored abstraction-driven interactive color palettes, showing that structured abstractions encourage exploratory design and improve perceptual metrics such as harmony and valence compared to traditional tools.

Palette-based harmonization methods have also gained attention. \cite{Xue2024} proposed a color naming–based harmonization approach that maps extracted palettes to perceptual prototypes, enabling visually coherent recoloring. These interactive and application-specific methods enhance usability but typically rely on predefined algorithms and color sets, limiting adaptability to complex or unseen color distributions.

Fuzzy-based approaches provide an alternative by explicitly modeling uncertainty in color perception. \cite{KimLee2024_FuzzyColorModel_2, 7044882, Shamoi2023} proposed a fuzzy color model to handle ambiguous color boundaries, while \cite{Zhou2021CCKMeans} introduced a CC\_K-Means algorithm that dynamically adjusts cluster numbers for clothing color matching. Although these methods improve flexibility, they often depend on handcrafted rules or application-specific assumptions.

Additional foundational work on cognitive and linguistic aspects of color perception is discussed in \cite{doi:10.1073/pnas.2400273121,DEREFELDT199569,Damasio1992}.


Overall, prior work shows that while clustering-based methods are effective for dominant color extraction, they often fail to align with human perceptual judgments because they rely on pixel-level statistics and fixed color representations \cite{11139272}. Perceptual and fuzzy models improve alignment, but are rarely grounded in large-scale human experiments. Interactive and application-driven approaches improve usability but lack generalizable perceptual foundations. These limitations motivate the need for a human-derived, experimentally grounded color representation that can be seamlessly integrated into existing clustering frameworks—an objective addressed in this work.


\section{Methods}

\subsection{Data Collection}

We collected the dataset by parsing several designer websites. Each image includes a photo with a professionally designed palette, and the hex codes for that palette. In total, 4562 images were obtained, and for each image, a 5-color palette was extracted. We use this dataset as ground truth because the palettes were manually selected by human experts rather than generated algorithmically.

Examples of extracted color palettes are shown in Table \ref{tab:palettes}.

\begin{table}[tb]
\centering
\small
\setlength{\tabcolsep}{3pt}
\renewcommand{\arraystretch}{1} 
\begin{tabular}{l c c c c c}
\hline
ID & \multicolumn{5}{c}{Color palette} \\
\hline

\multirow{2}{*}{1461}
 & \#070606 & \#959273 & \#CCD8B2 & \#FBFBF8 & \#6A8680 \\
 & \cellcolor[HTML]{070606}\color[HTML]{070606}\rule{0.5cm}{0.3cm}
 & \cellcolor[HTML]{959273}\color[HTML]{959273}\rule{0.5cm}{0.3cm}
 & \cellcolor[HTML]{CCD8B2}\color[HTML]{CCD8B2}\rule{0.5cm}{0.3cm}
 & \cellcolor[HTML]{FBFBF8}\color[HTML]{FBFBF8}\rule{0.5cm}{0.3cm}
 & \cellcolor[HTML]{6A8680}\color[HTML]{6A8680}\rule{0.5cm}{0.3cm} \\
\hline

\multirow{2}{*}{1462}
 & \#00747F & \#00B0BB & \#FFFFB0 & \#E9DFE0 & \#50525B \\
 & \cellcolor[HTML]{00747F}\color[HTML]{00747F}\rule{0.5cm}{0.3cm}
 & \cellcolor[HTML]{00B0BB}\color[HTML]{00B0BB}\rule{0.5cm}{0.3cm}
 & \cellcolor[HTML]{FFFFB0}\color[HTML]{FFFFB0}\rule{0.5cm}{0.3cm}
 & \cellcolor[HTML]{E9DFE0}\color[HTML]{E9DFE0}\rule{0.5cm}{0.3cm}
 & \cellcolor[HTML]{50525B}\color[HTML]{50525B}\rule{0.5cm}{0.3cm} \\
\hline

\multirow{2}{*}{1463}
 & \#4D3474 & \#8D6FBE & \#CEC3EB & \#EDEAF8 & \#BE5CD7 \\
 & \cellcolor[HTML]{4D3474}\color[HTML]{4D3474}\rule{0.5cm}{0.3cm}
 & \cellcolor[HTML]{8D6FBE}\color[HTML]{8D6FBE}\rule{0.5cm}{0.3cm}
 & \cellcolor[HTML]{CEC3EB}\color[HTML]{CEC3EB}\rule{0.5cm}{0.3cm}
 & \cellcolor[HTML]{EDEAF8}\color[HTML]{EDEAF8}\rule{0.5cm}{0.3cm}
 & \cellcolor[HTML]{BE5CD7}\color[HTML]{BE5CD7}\rule{0.5cm}{0.3cm} \\
\hline

\multirow{2}{*}{1464}
 & \#A8D1E8 & \#B3DBDA & \#FEE5E0 & \#FFBFC3 & \#A9A199 \\
 & \cellcolor[HTML]{A8D1E8}\color[HTML]{A8D1E8}\rule{0.5cm}{0.3cm}
 & \cellcolor[HTML]{B3DBDA}\color[HTML]{B3DBDA}\rule{0.5cm}{0.3cm}
 & \cellcolor[HTML]{FEE5E0}\color[HTML]{FEE5E0}\rule{0.5cm}{0.3cm}
 & \cellcolor[HTML]{FFBFC3}\color[HTML]{FFBFC3}\rule{0.5cm}{0.3cm}
 & \cellcolor[HTML]{A9A199}\color[HTML]{A9A199}\rule{0.5cm}{0.3cm} \\
\hline

\multirow{2}{*}{1465}
 & \#F8E905 & \#705183 & \#957C8F & \#A4DDF6 & \#24BEE8 \\
 & \cellcolor[HTML]{F8E905}\color[HTML]{F8E905}\rule{0.5cm}{0.3cm}
 & \cellcolor[HTML]{705183}\color[HTML]{705183}\rule{0.5cm}{0.3cm}
 & \cellcolor[HTML]{957C8F}\color[HTML]{957C8F}\rule{0.5cm}{0.3cm}
 & \cellcolor[HTML]{A4DDF6}\color[HTML]{A4DDF6}\rule{0.5cm}{0.3cm}
 & \cellcolor[HTML]{24BEE8}\color[HTML]{24BEE8}\rule{0.5cm}{0.3cm} \\
\hline

\end{tabular}
\caption{Examples of extracted color palettes.}
\label{tab:palettes}
\end{table}


\subsection{K-means Color Clustering}

K-means clustering, a widely used unsupervised learning technique \cite{jumb2014color, aruzhan, 10759889_11}. The algorithm partitions pixels into clusters based on color similarity, effectively summarizing the image’s chromatic structure.  The procedure consists of the following steps:

\begin{itemize}
    \item \textbf{Preprocessing.} Each image is transformed into a set of pixel vectors within a three-dimensional RGB color space.
    
    \item \textbf{Initialization.} The desired number of clusters $k$, corresponding to the number of dominant colors, is defined. Initial centroids are randomly positioned within the image color space.
    
    \item \textbf{Clustering.} Every pixel is assigned to the nearest centroid according to the Euclidean distance in RGB space:
    \begin{equation}
    \small
        d(c_1, c_2) = \sqrt{(r_2 - r_1)^2 + (g_2 - g_1)^2 + (b_2 - b_1)^2},
        \label{eq:rgbDistance}
    \end{equation}
    where $r$, $g$, and $b$ denote the red, green, and blue components of the colors, respectively.
    
    \item \textbf{Centroid Update.} Each centroid is recalculated as the mean of all pixels assigned to it, updating the RGB components to represent the cluster’s new center.
    
    \item \textbf{Iteration.} The assignment and update steps are repeated until convergence, when centroid positions stabilize, and dominant colors are effectively captured.
\end{itemize}

The algorithm outputs $k$ centroids representing the primary colors present in the image, serving as quantitative descriptors for subsequent visual and perceptual analysis. In this study, $k$ was set to five.

\subsection{RGB Histogram for Color Extraction}

A color histogram represents the distribution of distinct colors within a given color space in an image. It is widely used for statistical analysis, contrast enhancement \cite{KANG2011557}, color preservation \cite{BATZIOU2026105889}, and revealing features such as dominant colors. This procedure consists of the following steps:

\begin{itemize}
    \item \textbf{Preprocessing.} Each image is transformed into a set of pixel vectors within a three-dimensional RGB color space.

    \item \textbf{Computing.} Each dimension is divided into 8 equal bins. We go through each pixel and place it in a bin.

     \item \textbf{Sorting.} We sort bins by the number of pixels in them and retrieve the top $k$ bins.

     \item \textbf{Calculating.} For each bin, compute the RGB value as the center of the bin:

    \begin{equation}
    R = \left(r_{\text{bin}} + 0.5\right) \times \text{bin\_width}
    \label{eq:binRepresentative}
    \end{equation}

\end{itemize}

The algorithm outputs $k$ representatives of the most statistically frequent colors in an image. In this study, $k$ was set to five.

\subsection{CIELAB Color Space and CIEDE2000}

CIELAB is a three‑dimensional, approximately perceptually uniform color space \cite{10464290, muratbekova2025colormodelsimageprocessing}. It is defined by three axes: L* (lightness, 0 = black, 100 = white), a* (green to red), and b* (blue to yellow). While L* is bounded between 0 and 100, a* and b* have no fixed limits, though practical boundaries depend on the spectral properties of real-world objects and illuminants. 


CIELAB is widely used for objective color measurement across fields such as image processing \cite{Hassan2022A}, medicine \cite{Pérez2016Development}, food, and agriculture \cite{Salueña2019CIELAB}.

\begin{figure*}
    \centering
    \includegraphics[width=\linewidth]{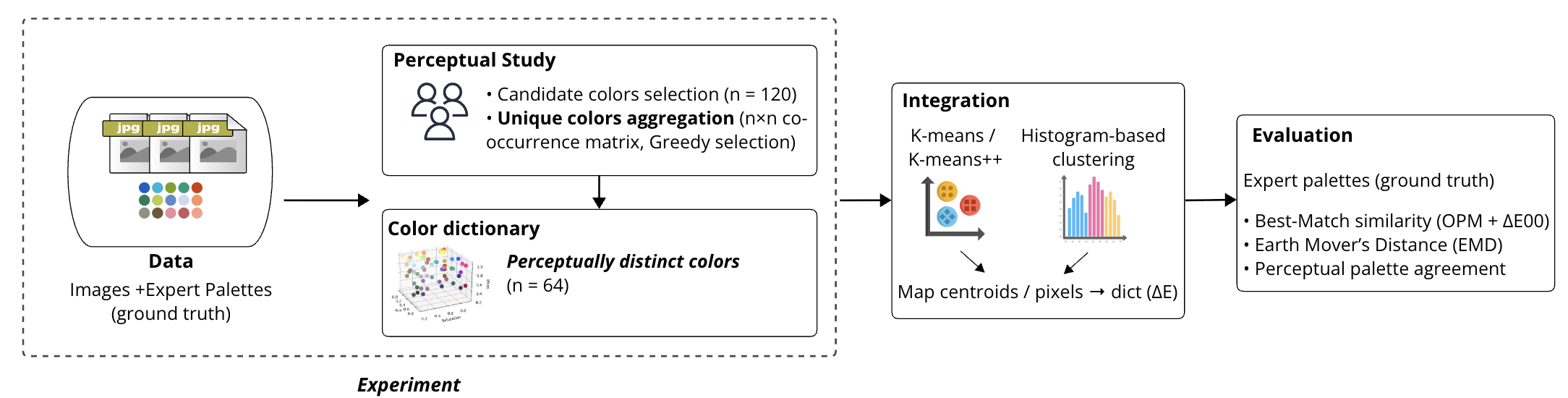}
    \caption{Two-level workflow: human-derived color dictionary integrated into dominant color extraction and evaluated against expert palettes}
    \label{fig:main}
\end{figure*}

The CIEDE2000 color difference ($\Delta E_{00}$) is a perceptually motivated metric designed to better align numerical color distances with human visual perception. $\Delta E_{00}$ is widely used in applications like color clustering, quality control, and image processing \cite{Gómez-Polo2020Comparison}. 
{
\small
\begin{equation}
\Delta E_{00} =
\sqrt{
\left( \frac{\Delta L'}{k_L S_L} \right)^2 +
\left( \frac{\Delta C'}{k_C S_C} \right)^2 +
\left( \frac{\Delta H'}{k_H S_H} \right)^2 +
R_T \left( \frac{\Delta C'}{k_C S_C} \right)
\left( \frac{\Delta H'}{k_H S_H} \right)
}
\label{eq:ciede2000}
\end{equation}
}

where $\Delta L'$, $\Delta C'$, and $\Delta H'$ denote the differences in lightness, chroma, and hue, respectively; $S_L$, $S_C$, and $S_H$ are weighting functions; $k_L$, $k_C$, and $k_H$ are parametric factors (set to 1 in this study); and $R_T$ is the rotation term accounting for the interaction between chroma and hue differences.

In this study, $\Delta E_{00}$ is used as the primary distance measure to ensure that color comparisons and mappings align with human perceptual judgments.

\subsection{Proposed Approach}



The proposed approach is illustrated in Fig. \ref{fig:main}. The workflow starts with a perceptual study in which participants select perceptually unique colors from an initial set, yielding a human-derived color dictionary (HCD) constructed from a co-occurrence matrix and greedy selection. Next, HCD is integrated into k-means clustering and histogram-based methods by perceptual mapping and is evaluated against expert palettes using Best-Match similarity and Earth Mover’s Distance to assess perceptual agreement.

We modified k-means by integrating HCD into the step of initializing centroids before running the algorithm. First, centroids were picked as usual: by randomly selecting five pixels from the image and using the advanced k-means++ algorithm. These five centroids are then mapped to the closest color in our human-based color dictionary (distance is calculated using the CIEDE2000 difference). In cases of repeated colors, duplicates were removed. To avoid the K decreasing, new centroids are added by selecting a random pixel from the original image. New centroids are also mapped; this process continues until the number of distinct centroids reaches five. Initial centroids are then passed down to the regular k-means function, where on each iteration, pixels are assigned to the closest centroid, and the new centroid of the cluster is recalculated using the mean of the whole cluster. We run an algorithm using both RGB and LAB color spaces.  

Our approach to color histograms was to first preprocess the original image. Initially, the image was resized to a smaller size, then each pixel was mapped to its closest color (distance is calculated by CIEDE2000 difference) from our derived human-based color dictionary. The edited image then undergoes the color histogram-based dominant color extraction (where each attribute R, G, B were divided into 8 bins) to obtain a palette of $n$ colors. In our case, $n$ was equal to 5.

\section{Experiment Design}


\begin{figure}[tb]
    \centering
    \begin{subfigure}{0.48\linewidth}
        \centering
        \includegraphics[width=\linewidth]{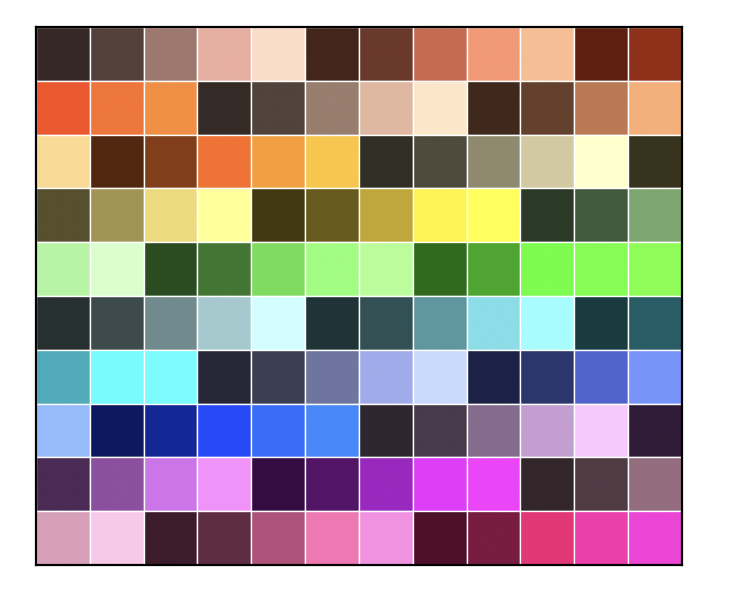}
        \caption{Dataset of Colors used in the experiment\cite{10464290}}
        \label{fig:dataset_a}
    \end{subfigure}
    \hfill
    \begin{subfigure}{0.48\linewidth}
        \centering
        \includegraphics[width=\linewidth]{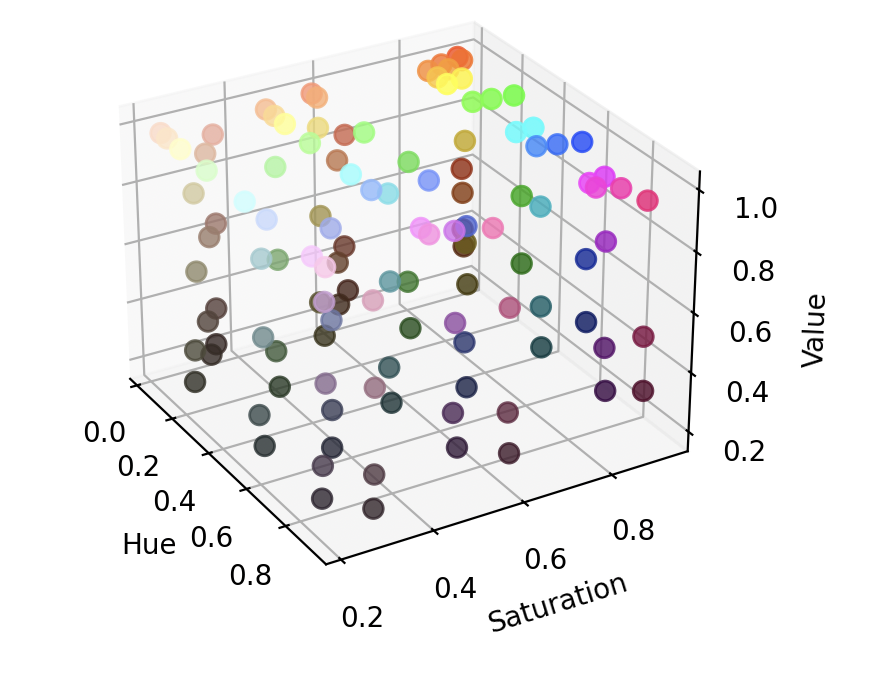}
        \caption{Dataset of Colors in HSV space}
        \label{fig:dataset_b}
    \end{subfigure}
    \caption{Color stimuli used in the human perception experiment}
    \label{fig:datasets}
\end{figure}

To determine the difference between colors by participants, this experiment presents 120 distinct colors from the dataset 
\cite{10464290}. Each entry in the dataset represents a unique color, although some of them are similar. Fig. \ref{fig:dataset_a} shows all 120 colors, while Fig. \ref{fig:dataset_b} illustrates their distribution in the HSV color space.
Participants will attempt to perceive the differences between these colors.

\begin{figure}[tb]
    \centering
    \includegraphics[width=\linewidth]{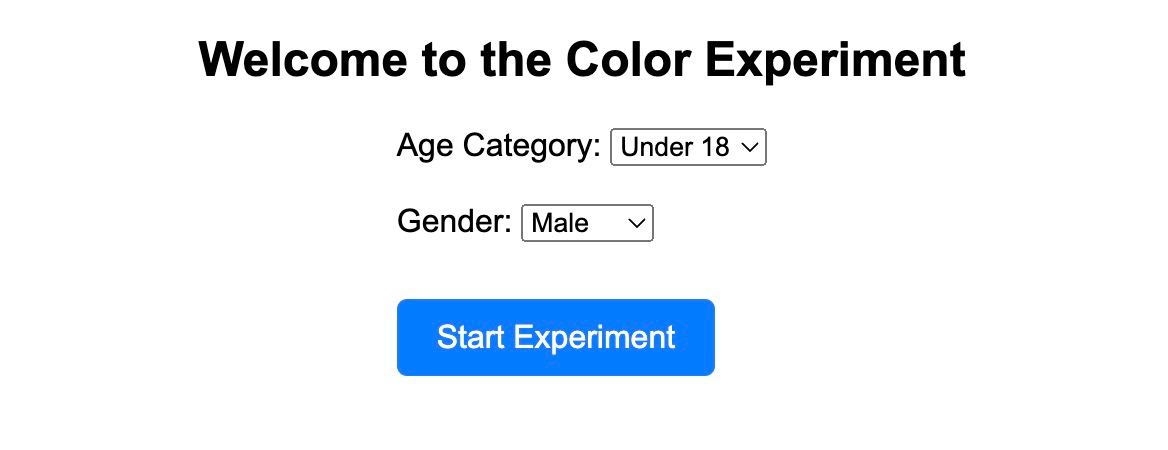}
    \caption{Start page of the experiment form}
    \label{fig:startpage}
\end{figure}

\begin{figure}[tb]
    \centering
    \includegraphics[width=\linewidth]{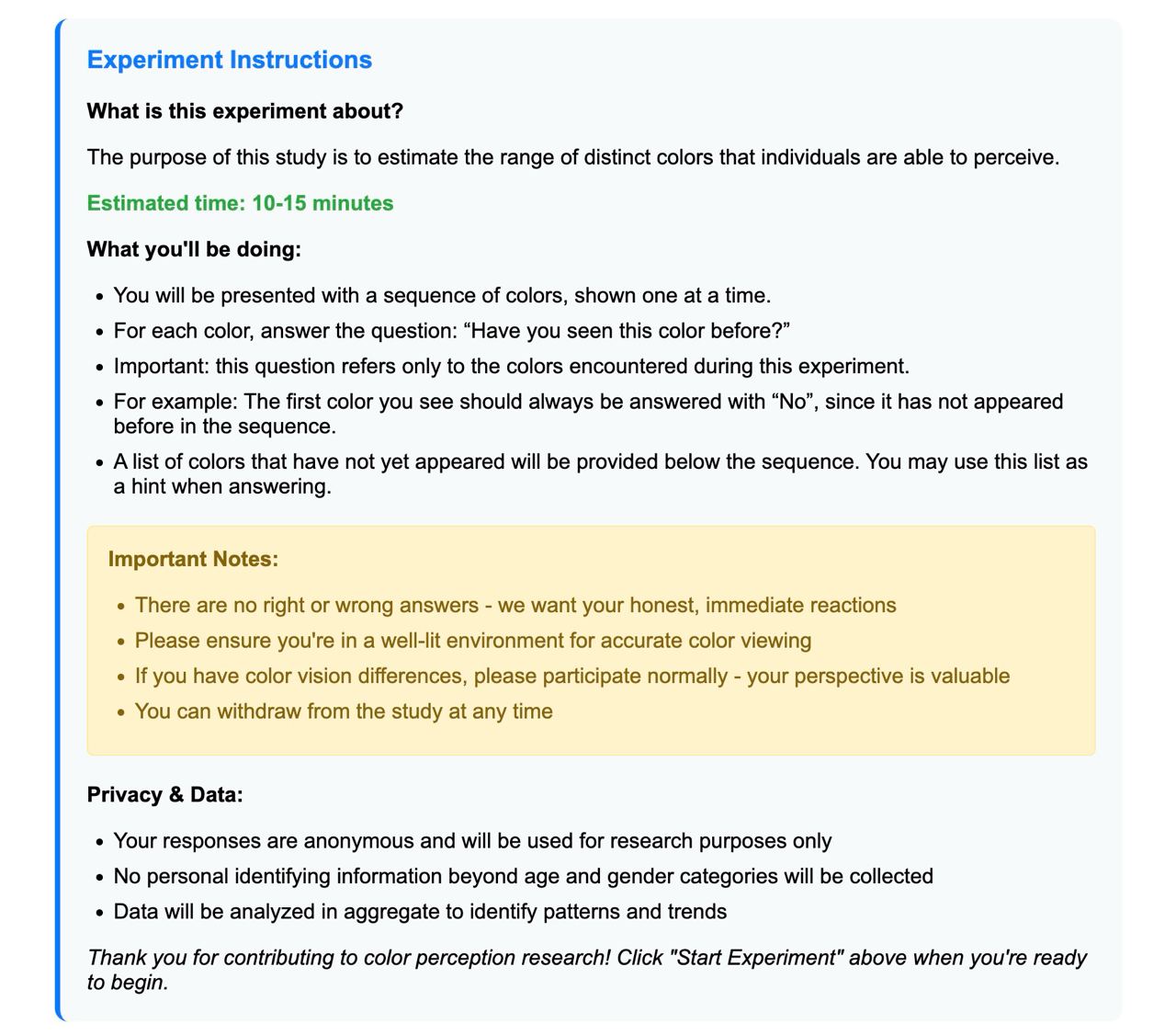}
    \caption{Instructions of the experiment}
    \label{fig:page}
\end{figure}

The tools for experiment design are based on a Python/Django project, using vanilla HTML/CSS/JavaScript for the UI concept, keeping only the necessary tools for the experiment, such as pandas and built-in libraries. The introduction page includes the "Age category" and "Gender" options, as well as the "Start" button (see Fig. \ref{fig:startpage}). It also includes "Experiment Instructions" with notes such as what the experiment aims to achieve, an approximate time for the participant, what they will be doing, and, of course, information about privacy and data (see Fig. \ref{fig:page}).

Once the experiment begins, the participant is presented with a 200×200 px square of color and asked, “Have you seen this color before?”, as illustrated in Fig. \ref{fig:main}. If the participant answers “No,” the color is added sequentially below as a 30×30 px color box, which can later be used for recognition and comparison with subsequent colors. If the participant answers “Yes,” no additional small color box is displayed.

The experiment ends when the participant has viewed all 120 distinct colors from the dataset, covering different color hues. 


\begin{figure}[h!]
    \centering
    \includegraphics[width=\linewidth]{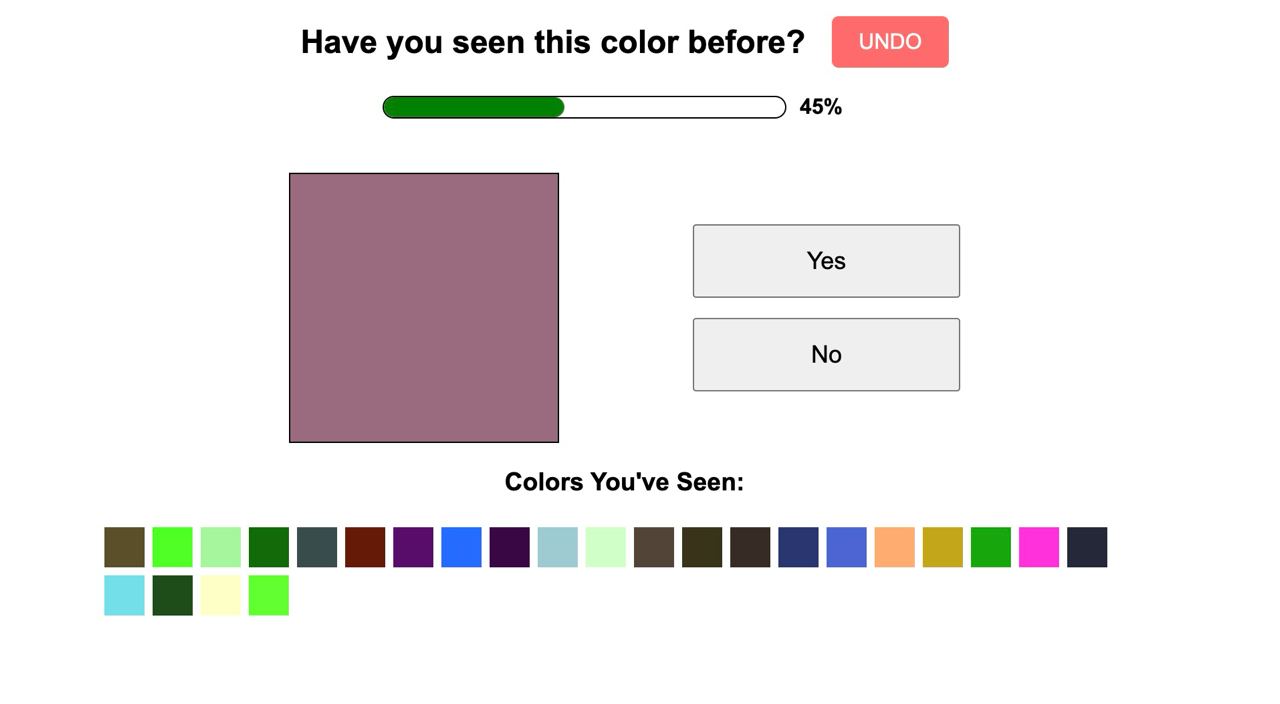}
    \caption{Main page of experiment}
    \label{fig:main}
\end{figure}

\section{Results}

 In this experiment, 110 participants were involved, including 59 females and 51 males. The age categories for both groups were under 18, 18–25, 26–35, 36–50, and over 50 years. The largest group consisted of participants aged 18–25 years (90 participants in total: 46 females and 44 males), representing 81.8\% of all participants, while the remaining age groups together accounted for 20 participants (19.2\%).

\subsection{Experiment Results}

\subsubsection{Human Color Dictionary}
According to experimental results, the average number of distinct colors is 64, accounting for 53.5\% of the total color stimuli presented.

\begin{figure}[h!]
    \centering
    \includegraphics[width=\linewidth]{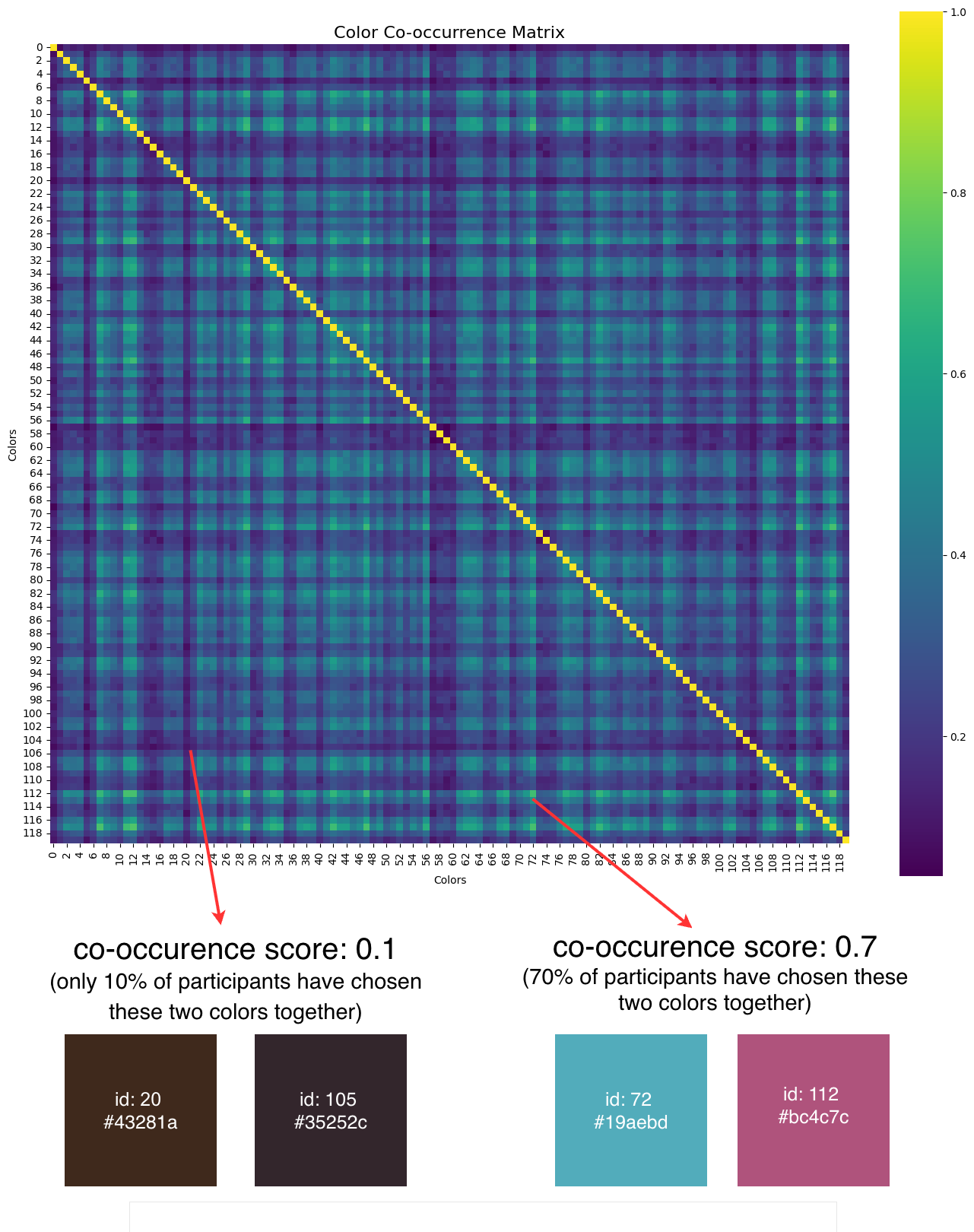}
    \caption{120 x 120 co-occurrence matrix and examples of color pairs.}
    \label{fig:co-occurrence-matrix}
\end{figure}

\begin{figure}[tb]
    \centering
    \begin{subfigure}{0.48\linewidth}
        \centering
        \includegraphics[width=\linewidth]{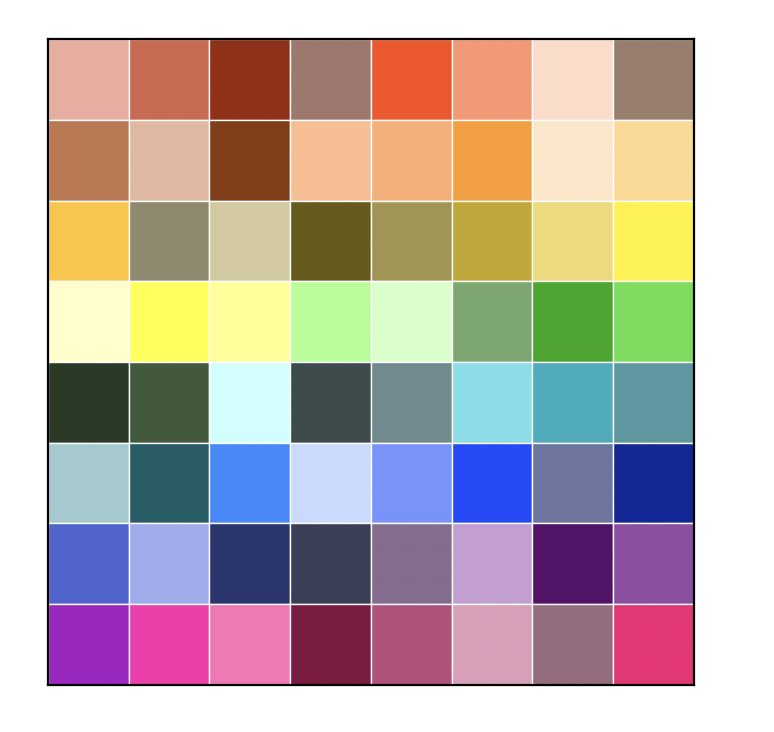}
        \caption{Derived colors as a grid}
        \label{fig:derived_dict_a}
    \end{subfigure}
    \hfill
    \begin{subfigure}{0.48\linewidth}
        \centering
        \includegraphics[width=\linewidth]{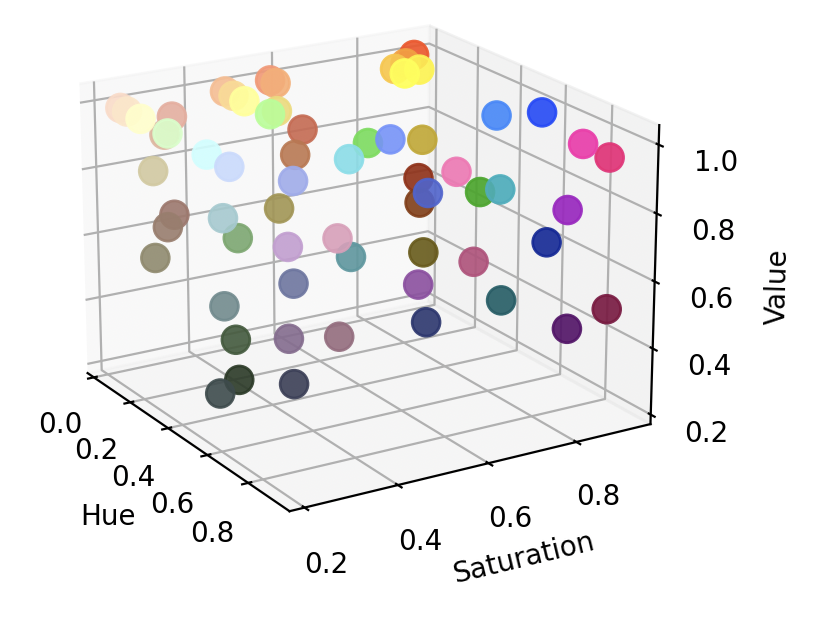}
        \caption{ Derived Colors in HSV space}
        \label{fig:derived_dict_b}
    \end{subfigure}
    \caption{Human color dictionary containing perceptually distinct colors retrieved from the experiment}
    \label{fig:derived_dict}
\end{figure}

Based on the results of the experiment, a 120 × 120 co-occurrence matrix (Fig.~\ref{fig:co-occurrence-matrix}) was constructed, where each cell represents how often the given pair of colors (row and column) was marked as distinct by the same participant. Using this matrix, a greedy selection algorithm was applied. First, the most distinctive color (the one marked as unique by most participants) is selected and added to the dictionary. Next, the color that most frequently co-occurred with the previous one is selected. This process was repeated until we derived a human-based color dictionary with 64 perceptually distinct colors. Derived colors and their distribution in HSV space are shown in Fig.~\ref{fig:derived_dict}.

\subsubsection{Examples}

\begin{figure*}[h!]
    \centering
    \includegraphics[width=\linewidth]{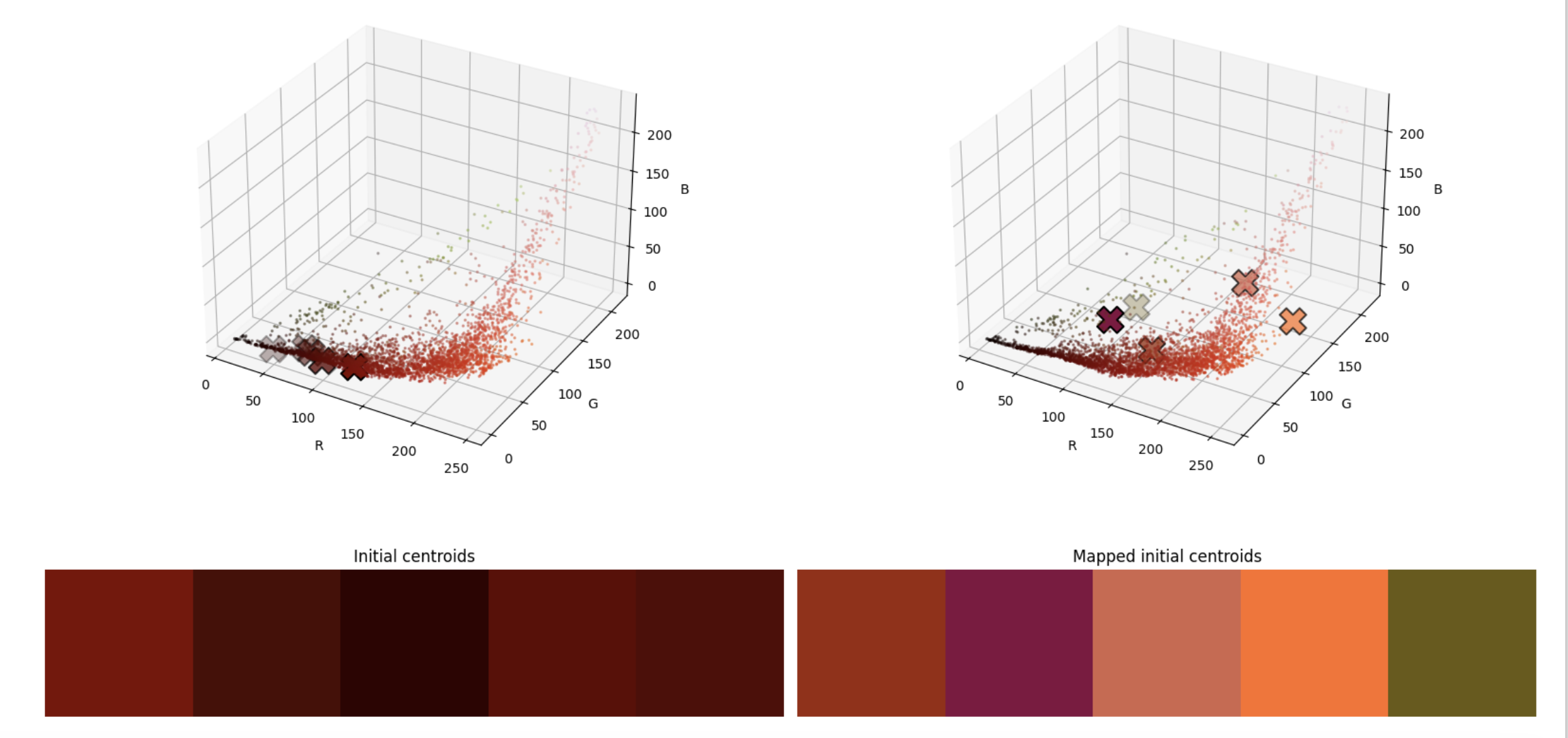}
    \caption{Initial centroids distribution in RGB space before and after proposed manipulations.}
    \label{fig:centroid_3d}
\end{figure*}

\begin{figure}[h!]
    \centering
    \includegraphics[width=1\linewidth]{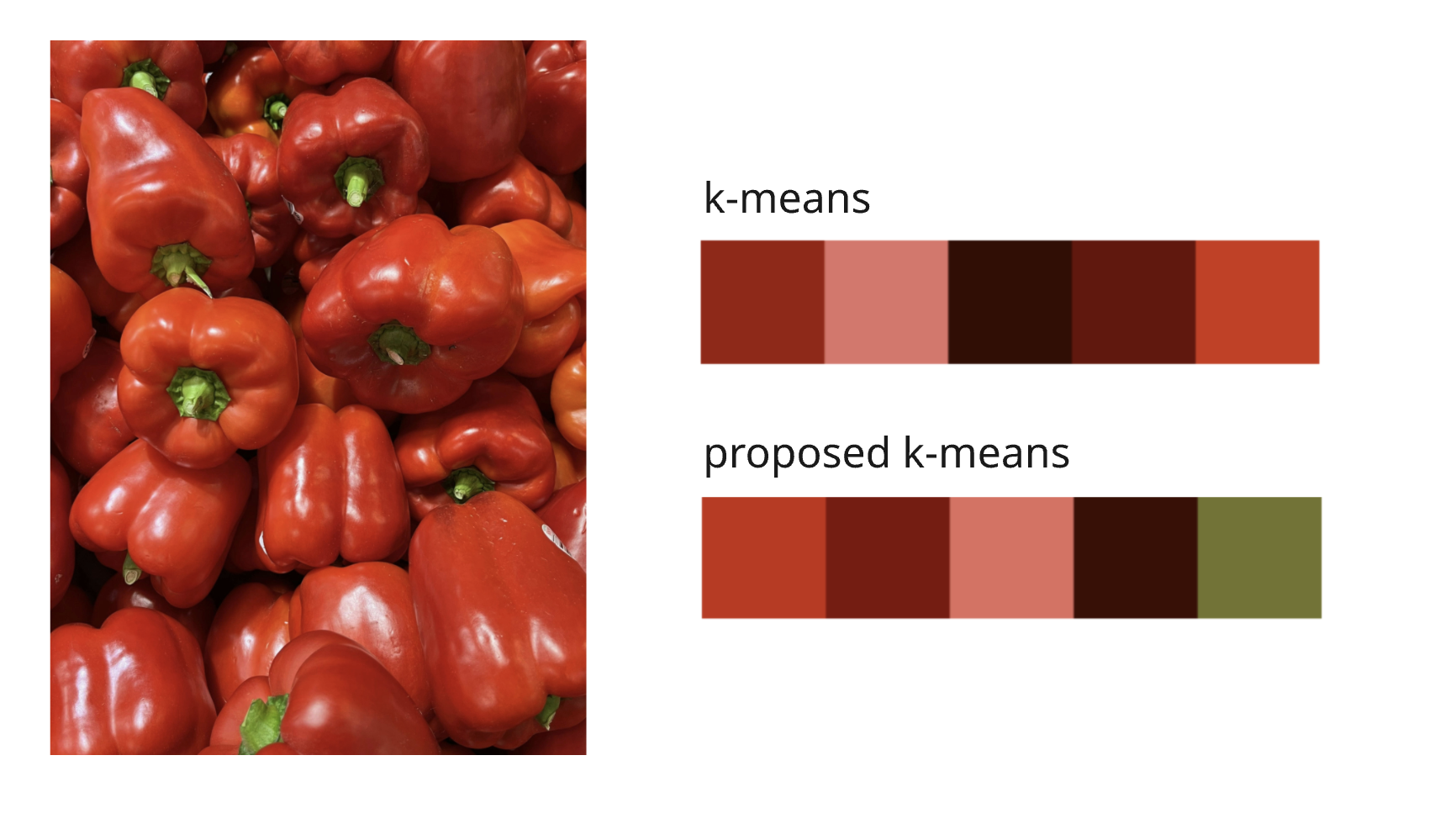}
    \caption{Comparison of the results of the original k-means and proposed k-means in RGB space.}
    \label{fig:pepper_kmeans_example}
\end{figure}

 Table~\ref{table:mapping_process} illustrates the step-by-step process of how initial centroids (that were chosen randomly) for the k-means algorithm were modified with our approach. This results in more perceptually evenly distributed centroids as shown in Fig.~\ref{fig:centroid_3d}; in this case, the distribution is in RGB space. The final side-by-side comparison of the classic k-means method and the proposed k-means in RGB space is shown in Fig.~\ref{fig:pepper_kmeans_example}.

\begin{table}[h]
\centering
\small
\setlength{\tabcolsep}{6pt} 
\renewcommand{\arraystretch}{1.4}
\begin{tabular}{l c c c c c}
\hline
\multicolumn{6}{l}{\textbf{Process Stage \& Centroid Evolution}} \\
\hline

\multicolumn{6}{l}{\textit{Initializing centroids}} \\
& \#7D0A01 & \#4B0B05 & \#300201 & \#5F0602 & \#520805 \\
& \cellcolor[HTML]{7D0A01}\color[HTML]{7D0A01}\rule{0.6cm}{0.3cm}
& \cellcolor[HTML]{4B0B05}\color[HTML]{4B0B05}\rule{0.6cm}{0.3cm}
& \cellcolor[HTML]{300201}\color[HTML]{300201}\rule{0.6cm}{0.3cm}
& \cellcolor[HTML]{5F0602}\color[HTML]{5F0602}\rule{0.6cm}{0.3cm}
& \cellcolor[HTML]{520805}\color[HTML]{520805}\rule{0.6cm}{0.3cm} \\ \hline

\multicolumn{6}{l}{\textit{Mapping to color dictionary}} \\
& \#9B290D & \#830E40 & \#830E40 & \#9B290D & \#9B290D \\
& \cellcolor[HTML]{9B290D}\color[HTML]{9B290D}\rule{0.6cm}{0.3cm}
& \cellcolor[HTML]{830E40}\color[HTML]{830E40}\rule{0.6cm}{0.3cm}
& \cellcolor[HTML]{830E40}\color[HTML]{830E40}\rule{0.6cm}{0.3cm}
& \cellcolor[HTML]{9B290D}\color[HTML]{9B290D}\rule{0.6cm}{0.3cm}
& \cellcolor[HTML]{9B290D}\color[HTML]{9B290D}\rule{0.6cm}{0.3cm} \\ \hline

\multicolumn{6}{l}{\textit{Duplicate removal}} \\
& \#9B290D & \#830E40 & -- & -- & -- \\
& \cellcolor[HTML]{9B290D}\color[HTML]{9B290D}\rule{0.6cm}{0.3cm}
& \cellcolor[HTML]{830E40}\color[HTML]{830E40}\rule{0.6cm}{0.3cm}
& \cellcolor[HTML]{FFFFFF}\color[HTML]{FFFFFF}\rule{0.6cm}{0.3cm} 
& \cellcolor[HTML]{FFFFFF}\color[HTML]{FFFFFF}\rule{0.6cm}{0.3cm}  
& \cellcolor[HTML]{FFFFFF}\color[HTML]{FFFFFF}\rule{0.6cm}{0.3cm}  \\ \hline

\multicolumn{6}{l}{\textit{Adding new colors }} \\
& \#9B290D & \#830E40 & \#D3654C & \#FF6D25 & \#6A590E \\
& \cellcolor[HTML]{9B290D}\color[HTML]{9B290D}\rule{0.6cm}{0.3cm}
& \cellcolor[HTML]{830E40}\color[HTML]{830E40}\rule{0.6cm}{0.3cm}
& \cellcolor[HTML]{D3654C}\color[HTML]{D3654C}\rule{0.6cm}{0.3cm}
& \cellcolor[HTML]{FF6D25}\color[HTML]{FF6D25}\rule{0.6cm}{0.3cm}
& \cellcolor[HTML]{6A590E}\color[HTML]{6A590E}\rule{0.6cm}{0.3cm} \\ 
\hline

\end{tabular}
\caption{Step-by-step mapping and refinement of initial centroids into the final color palette.}
\label{table:mapping_process}
\end{table}

\begin{figure}[h]
    \centering
    \begin{subfigure}{0.48\linewidth}
        \centering
        \includegraphics[width=\linewidth]{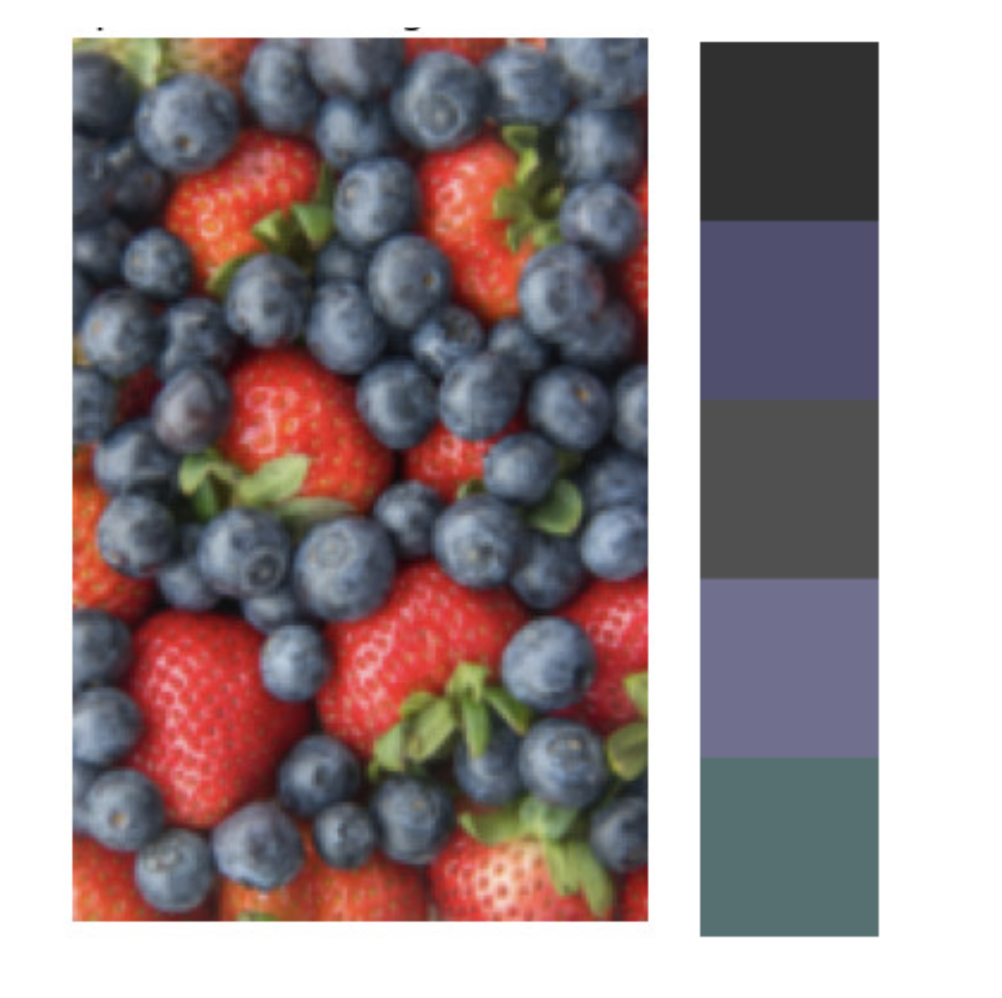}
        \caption{Original image and extracted dominant colors}
        \label{fig:berries_a}
    \end{subfigure}
    \hfill
    \begin{subfigure}{0.48\linewidth}
        \centering
        \includegraphics[width=\linewidth]{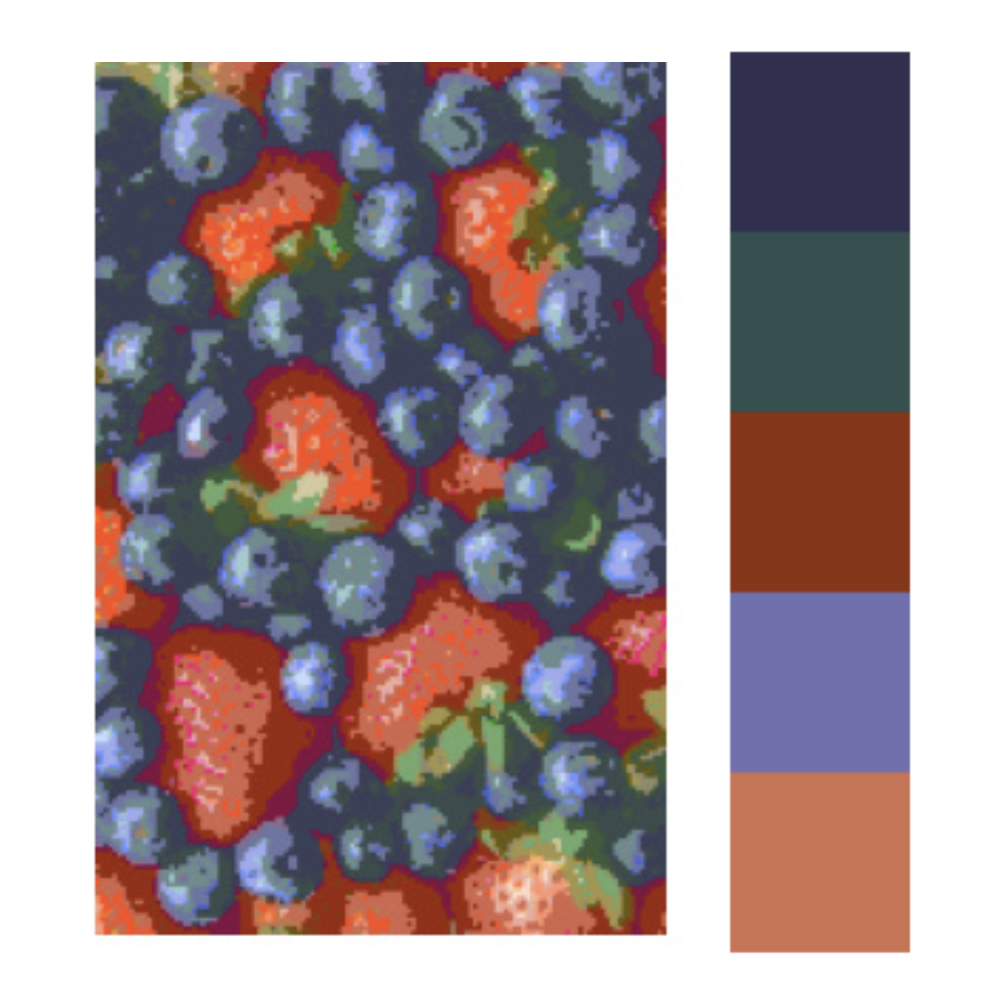}
        \caption{Image mapped to proposed HCD and extracted dominant colors}
        \label{fig:berries_b}
    \end{subfigure}
    \caption{Histogram methods comparison}
    \label{fig:berries}
\end{figure}
Fig.~\ref{fig:berries} compares the results of the color histogram algorithm and our proposed approach. First, the image was resized, and the palette was extracted using color histograms (Fig. \ref{fig:berries_a}). Afterwards, each pixel of that image was mapped to its closest color from our human-based color dictionary, and a palette was extracted from the edited image (Figure \ref{fig:berries_b}).



\subsubsection{Performance Evaluation}
To assess the performance of our models, we calculated how similar the obtained palettes were to the "ground-truth" ones. Similarity scores were calculated using 2 methods: Earth Mover's Distance (EMD) and the Permutation-Invariant Palette Similarity Method, which provides Optimal Palette Matching (OPM).

EMD compares two distributions by quantifying the cost of transforming one distribution into the other. It tends to be more comprehensive when comparing color palettes because of its consistently defined lower and upper bounds \cite{Weller2019ColorDistance}.

OPM goes through all possible permutations of the evaluated palette to find the most optimal (best) similarity score. For each permutation $\pi \in \Pi(P_2)$, pairwise CIEDE2000 distances are computed between corresponding colors of the first palette $P_1$ and the permutation $\pi$ of the second palette $P_2$ (assuming both palettes have the same length). The similarity for a single permutation is calculated as
\begin{equation}
S(\pi) = 100 \cdot \min \Bigg( clip \Big( 1 - \frac{\Delta E_{00}(P_1, \pi)}{100}, \, 0, \, 1 \Big) \Bigg)
\end{equation}

where $\Delta E_{00}(P_1, \pi)$ is the vector of CIEDE2000 differences for each color pair, and the $\min$ operator selects the lowest similarity across all pairs. The best similarity score across all permutations is then

\begin{equation}
S_{\text{best}} = \max_{\pi \in \Pi(P_2)} S(\pi)
\end{equation}


During the comparison of palettes, we have normalized both scores and used them as percentages. 
During the performance evaluation, 4562 photos were processed. We extracted dominant colors using base models: histogram, RGB k-means, LAB k-means, RGB k-means++, and LAB k-means++; as well as our customized methods, in which we applied mapping to a human-based color dictionary to enhance results. As a result, we made the next improvements: 


\begin{table}[tb]
\centering
\small
\renewcommand{\arraystretch}{1.1}
\begin{tabular}{l cc}
\hline
Method / Configuration & Best-Match $\uparrow$ & EMD $\uparrow$ \\
                       & [\%]                  & [\%]           \\
\hline

\multicolumn{3}{l}{\textit{Histogram-based baselines}} \\
Color histogram (RGB) & 64.83 & 92.04 \\
Color histogram + mapped to dictionary & \textbf{70.78} & \textbf{93.37} \\[6pt]

\multicolumn{3}{l}{\textit{K-means, Space = Lab}} \\
Random + None + Lab (Original) & 77.50 & 95.13 \\
Random + Random\_replace + Lab & \textbf{77.92} & \textbf{95.16} \\[6pt]

\multicolumn{3}{l}{\textit{K-means, Space = RGB}} \\
Random + None + RGB (Original) & 75.88 & 94.61 \\
Random + Random\_replace + RGB & \textbf{76.08} & \textbf{94.66} \\[6pt]

\multicolumn{3}{l}{\textit{K-means++, Space = RGB}} \\
Smart + None + RGB (Baseline) & 75.85 & 94.59 \\
Smart + Random\_replace + RGB & \textbf{76.08} & \textbf{94.65} \\
\hline
\end{tabular}
\caption{Comparative evaluation of dominant-color extraction variants against expert palettes using two similarity metrics: Best-Match and Earth Mover's Distance (EMD). Higher values indicate better agreement with the ground-truth palettes. Within each block, the best-performing configuration is highlighted in bold.}
\label{tab:comparative_results}
\end{table}

Table~\ref{tab:comparative_results} compares histogram and K-means variants with respect to palette agreement against expert-designed ground truth. Mapping histogram colors to the perceptual dictionary substantially improves Best-Match similarity (64.83 $\rightarrow$ 70.78) and also increases EMD. Across K-means settings, operating in CIE~Lab generally yields higher scores than RGB.

\subsubsection{Discussion}

\begin{figure}[tb]
    \centering
    \includegraphics[width=0.75\linewidth]{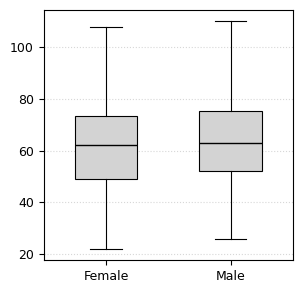}
    \caption{Boxplot for the number of colors marked as unique by males and females}
    \label{fig:boxplot}
\end{figure}

\textit{Gender and Color Perception.} As shown in Table~\ref{table:experiment_gender_statistics} and (Fig.~\ref{fig:boxplot}), the average number of distinguished colors by gender was 63.37 for females and 63.94 for males, respectively.  Based on these results, there’s no evidence of a meaningful gender difference in either the average number of unique colors perceived, their median, or the variability across participants. 
 
\begin{table}[tb]
\centering
\small
\setlength{\tabcolsep}{4pt} 
\renewcommand{\arraystretch}{1}
\begin{tabular}{l c c c c c} 
\hline
Color stimuli & Gender & Mean & Median & Std. Dev. \\ 
\hline
Marked as repeated & Female & 56.24 & 57.0 & 19.76 \\
Marked as repeated & Male   & 55.47 & 56.0 & 18.56 \\
Marked as unique & Female & 63.37 & 62.0 & 19.83 \\
Marked as unique & Male & 63.94 & 63.0 & 18.14 \\
\hline
\end{tabular}
\caption{Descriptive statistics of color recognition by gender. Mean, median, and standard deviation (Std. Dev.) are reported for females and males for repeated and distinguished colors.}
\label{table:experiment_gender_statistics}
\end{table}

\begin{figure}[tb]
    \centering
    \begin{subfigure}[b]{0.5\textwidth}
        \centering
        \includegraphics[width=\textwidth]{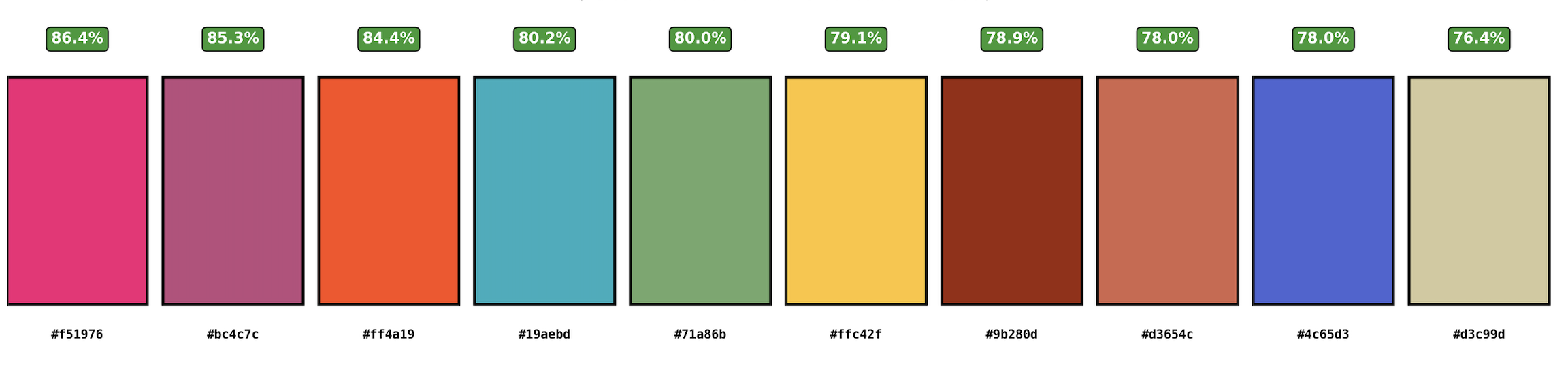}
        \caption{Top 10 Most Distinguished Colors by Human Perception}
        \label{fig:most_distinguishable}
    \end{subfigure}
    
    \vspace{1em} 

    \begin{subfigure}[b]{0.5\textwidth}
        \centering
        \includegraphics[width=\textwidth]{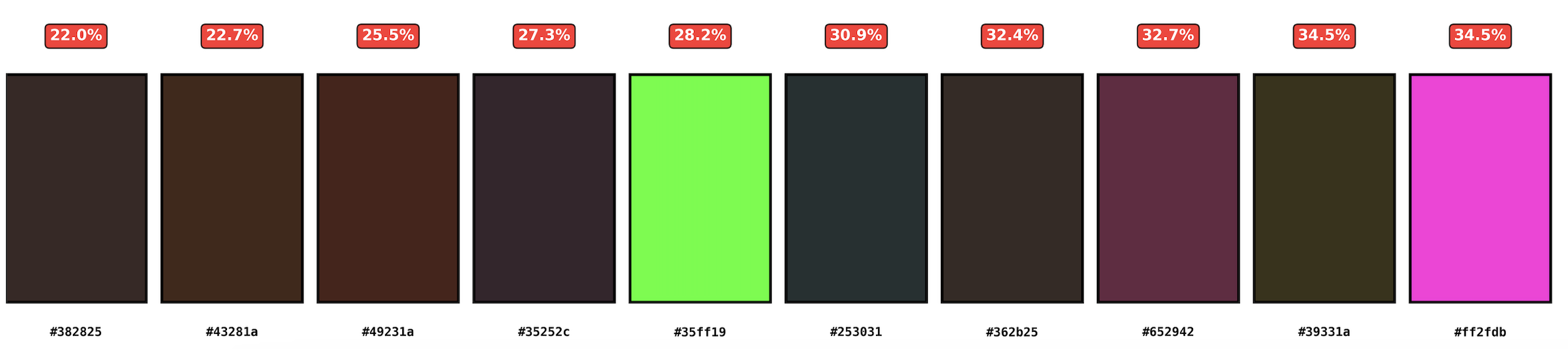}
        \caption{Top 10 Least Distinguished Colors by Human Perception}
        \label{fig:least_distinguishable}
    \end{subfigure}

    \caption{Human perception of color distinctiveness: (a) most distinguished colors, (b) least distinguished colors.}
    \label{fig:distinguishability_comparison}
\end{figure}

\begin{figure*}[h]
    \centering
    \includegraphics[width=1\textwidth]{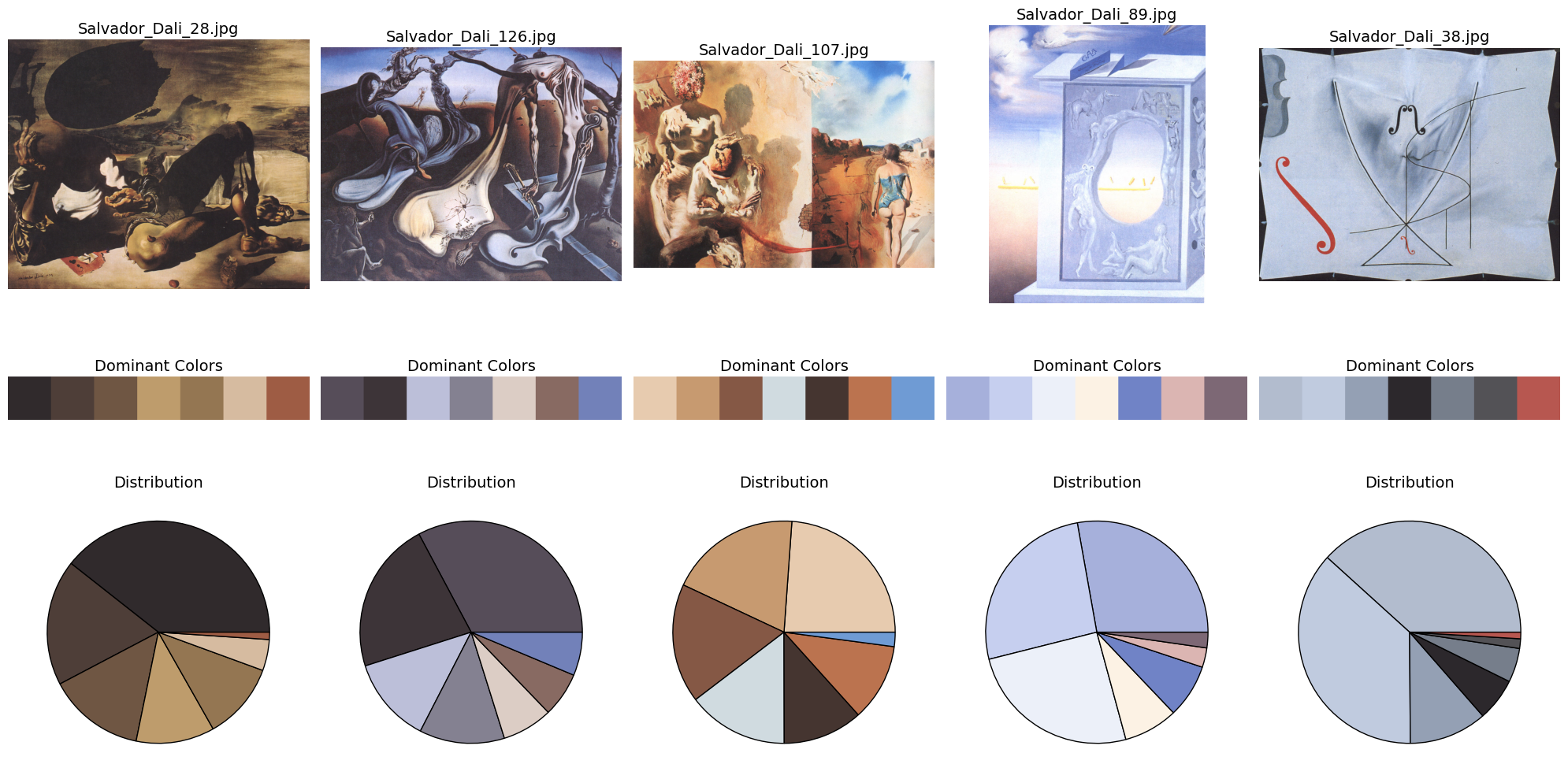}
    \caption{Example of palette extraction over Salvador Dali \cite{best-artworks} subset (n = 139); 5 images are randomly selected; extracted dominant color palettes with proportional distributions (pie charts).}
    \label{fig:palatte_example}
\end{figure*}

\begin{figure*}[h]
    \centering
    \includegraphics[width=1\textwidth]{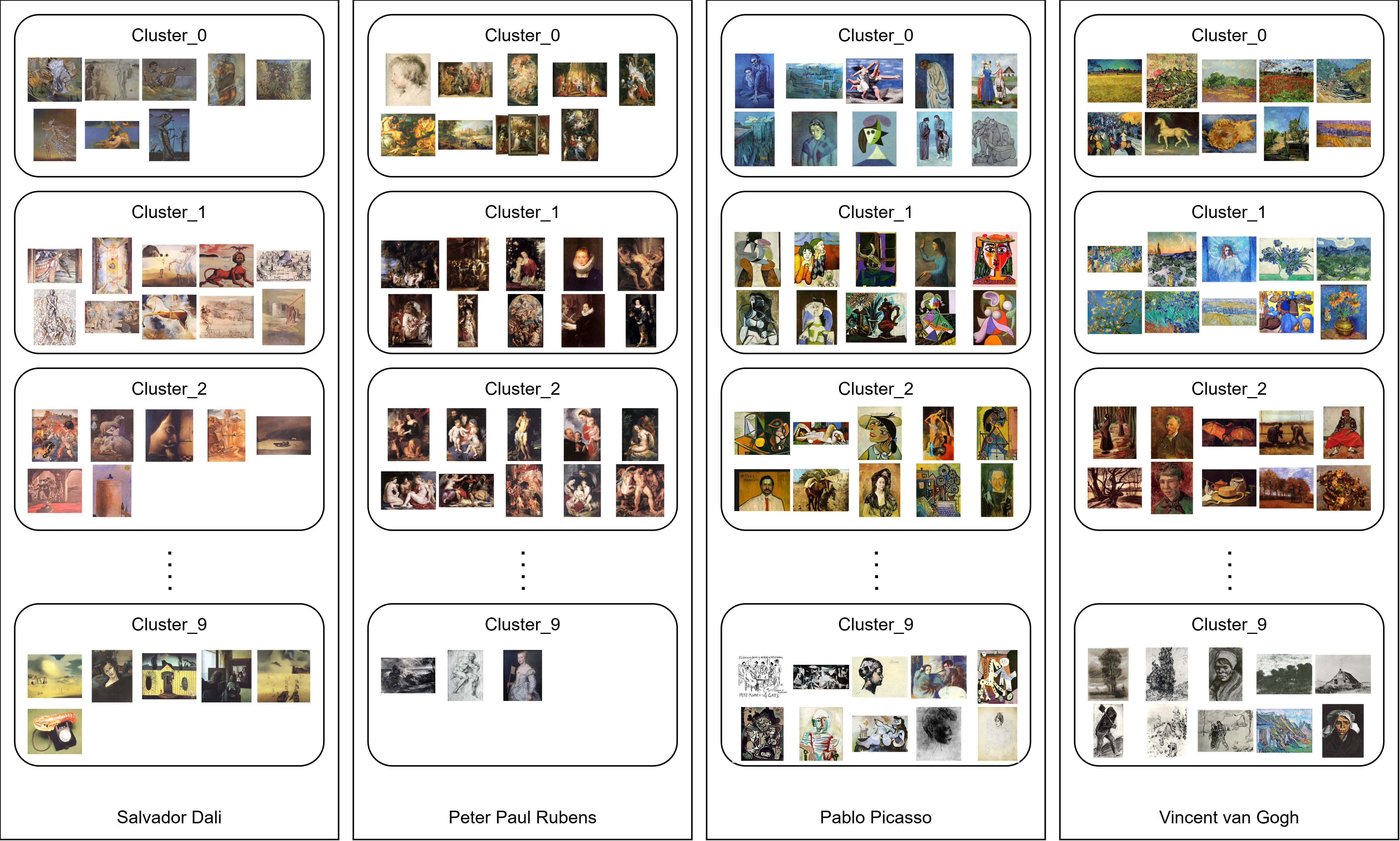}
    \caption{Paintings by S. Dali, P. Rubens, P. Picasso, and V. van Gogh \cite{best-artworks}. Example of palette-based clustering using an agglomerative (hierarchical) clustering algorithm with K=10 clusters. For visualization, up to 10 images are shown per cluster.}
    \label{fig:cluster_painters}
\end{figure*}

\begin{table}[tb]
\small
\centering
\begin{tabular}{l c c}
\hline
Attribute & Unique count & Ratio \\
\hline
\multicolumn{3}{c}{Saturation} \\
High    & 2391 & 0.342 \\
Low     & 2376 & 0.339 \\
Medium  & 2233 & 0.319 \\
\hline
\multicolumn{3}{c}{Intensity} \\
Medium  & 1774 & 0.253 \\
Pale    & 1496 & 0.214 \\
Light   & 1386 & 0.198 \\
Deep    & 1349 & 0.193 \\
Dark    & 995  & 0.142 \\
\hline
\end{tabular}
\caption{Distribution of saturation and intensity attributes among colors marked as unique in the experiment.}
\label{tab:saturation_intensity_not_seen}
\end{table}

\begin{table}[tb]
\small
\centering
\begin{tabular}{l l c c}
\hline
Saturation & Intensity & Unique count & Ratio \\
\hline
Medium & Medium & 604 & 0.086 \\
Low    & Medium & 589 & 0.084 \\
High   & Medium & 581 & 0.083 \\
High   & Deep   & 570 & 0.081 \\
Low    & Pale   & 548 & 0.078 \\
Low    & Light  & 516 & 0.074 \\
Medium & Pale   & 516 & 0.074 \\
Medium & Light  & 444 & 0.063 \\
High   & Pale   & 432 & 0.062 \\
High   & Light  & 426 & 0.061 \\
Low    & Deep   & 413 & 0.059 \\
High   & Dark   & 382 & 0.055 \\
Medium & Deep   & 366 & 0.052 \\
Low    & Dark   & 310 & 0.044 \\
Medium & Dark   & 303 & 0.043 \\
\hline
\end{tabular}
\caption{Joint distribution of saturation and intensity attributes among colors marked as unique in the experiment.}
\label{tab:sat_int_combination_not_seen}
\end{table}

 Gender does not appear to be an important driver of the number of unique colors participants identify (i.e., how much they vary from one another). High values for standard deviation (19.83 for females and 18.14 for males) show that any gender differences that exist here are small relative to the participant-to-participant variation.

In contrast, prior studies suggest that sex differences in color perception exist \cite{Abramov2012Sex, Pardo2007An}. Previous studies have reported sex-related differences in color perception, suggesting that females, on average, may exhibit higher sensitivity to subtle color differences and finer chromatic discrimination \cite{Gómez-Polo2016Comparison, Gómez-Polo2020Comparison}, particularly for specific hues such as red, green, blue, and orange \cite{Huang2016Gender}. Another study reports that females can perceive more color shades than males \cite{Jaint2010Gender}. These findings are often attributed to differences in socialization, genetic influences, or retinal physiology \cite{Bimler2002Sex}.

Our perceptual experiment does not show a meaningful gender difference in the number of uniquely perceived colors. This may be partly explained by the relatively homogeneous age range of the participants, who were predominantly young adults (18-25 years), which may limit the emergence of age- and gender-related perceptual differences observed in broader populations. The relatively large standard deviations (19.83 for females and 18.14 for males) indicate that inter-individual variability is considerably greater than any systematic gender-related effect. This suggests that while sex differences may emerge in fine color discrimination tasks, they do not substantially influence higher-level judgments of perceptual color uniqueness as investigated in this study.

\textit{Impact of Intensity and Saturation on Color Perception.} The Fig.~\ref{fig:most_distinguishable} shows that unique colors have high saturation or unique hues that stand out from the rest of the palette. Meanwhile, as shown in the Fig.~\ref{fig:least_distinguishable}, dark colors were most often perceived as indistinguishable by human perception from other colors in the set. Saturation was categorized as low, medium, or high, and intensity as light, pale, medium, deep, or dark, using the value ranges defined by fuzzy sets in the recent study \cite{10464290}.

Furthermore, Table \ref{tab:saturation_intensity_not_seen} illustrates that high-saturation and low-saturation colors are most frequently distinguished from colors with medium saturation.  This finding confirms previously reported results that, in chromatic discrimination tasks, higher color saturation generally makes differences easier to see \cite{Igarashi2022Chromatic}.

As for the intensity, medium intensity dominates. Dark and deep colors are the least distinguishable and likely collapse perceptually.

Next, Table \ref{tab:sat_int_combination_not_seen} shows a strong interaction effect between saturation and intensity on color distinguishability. Rather than either attribute alone, their combination determines how often colors are perceived as distinct. All three saturation levels reach their maximum distinguishability at medium intensity. Medium intensity provides optimal luminance contrast, allowing saturation differences to be perceptually resolved. Dark intensity yields the lowest distinguishability, regardless of saturation. This indicates that luminance contrast plays a dominant role in perceptual color separation, with saturation effects becoming secondary when intensity is suboptimal.






\subsection{Application Example}
 In practical applications, it is crucial to prioritize human perception over purely mathematical accuracy. Broad application areas include graphic and interior design, where improved dominant color extraction enables more accurate palette matching; art and color theory, through more reliable computational studies of composition, brightness, and shadows; and the fashion industry, through more harmonized color recommendations.  The method can also be used to generate datasets of perceptually grounded color palettes for training, fine-tuning, and evaluating neural networks \cite{article1}. Additionally, the approach can be applied to thermal-image colorization with more perceptually distinct pseudo-colors, making images easier to interpret \cite{article}. The method can be extended to construct culture- and language-specific color vocabularies \cite{doi:10.1073/pnas.2400273121,DEREFELDT199569,Damasio1992}.

One practical application is image clustering, where the set partitions are highly sensitive to the underlying color vocabulary used to represent each image. Since the proposed K-means with a human-based color vocabulary yields more perceptually accurate results in dominant color extraction, it can enhance overall quality across image-processing tasks. We evaluate the method on the open-source \textit{Best Artworks of All Times} dataset \cite{best-artworks}. 

For each image, K-means is performed in the Lab color space. Centroids are initialized via random sampling and then replaced with the nearest HCD samples using the CIEDE2000 difference. Duplicate centroids are removed and replenished via random pixel resampling. For visualization, 5 images are randomly selected from the 139-image subset, and their dominant colors, along with the corresponding percentage distributions after K-means-based color replacement, are shown in Fig.~\ref{fig:palatte_example}. To test the proposed method, we again used the same artworks dataset \cite{best-artworks} and clustered 4 painters into 10 clusters based on their extracted color features using an agglomerative (hierarchical) clustering algorithm (Fig.~\ref{fig:cluster_painters}). 

Human-based color set for dominant color retrieval provides a perceptually grounded clustering, resulting in groups of images with similar color appearance.




\section{Conclusion}

This paper presents a human-centered framework for dominant color extraction that integrates experimentally derived perceptual knowledge into traditional clustering and histogram-based methods. We constructed a Human Color Dictionary (HCD) that captures colors humans perceive as distinct, and embedded it in standard extraction pipelines via perceptual color mapping in the CIE Lab space using the CIEDE2000 metric.

Quantitative evaluation against expert-designed ground-truth palettes demonstrates that the proposed integration consistently improves performance across all tested configurations. As summarized in Table~\ref{tab:comparative_results}, mapping colors to the human-derived dictionary yields higher agreement with expert palettes than baseline approaches. These results confirm that incorporating perceptually grounded color representations enhances the effectiveness without altering the core structure of existing algorithms.

Our results show no meaningful gender differences in perceived color uniqueness; instead, color distinguishability is driven by the interaction of saturation and intensity, with highly saturated colors being most distinct and dark or deep colors tending to perceptually collapse.

Overall, the findings demonstrate that integrating a human-derived color dictionary into computational color extraction yields color extraction results that more closely reflect human visual perception than those from traditional clustering algorithms. Future research will explore cultural variations in color naming and extend the approach by integrating with deep learning–based color quantization techniques.






\bibliographystyle{elsarticle-harv} 
\bibliography{Bibliography}






\end{document}